\documentclass[%
reprint,
superscriptaddress,
amsmath,amssymb,
aps,
]{revtex4-2}

\usepackage{graphicx}% Include figure files
\usepackage{dcolumn}% Align table columns on decimal point
\usepackage{bm}% bold math

\usepackage{multirow}

\begin{document}

\preprint{APS/123-QED}

\title{Soft X-ray phase nano-microscopy of micrometre-thick magnets \\}% Force line breaks with \\
%\thanks{A footnote to the article title}%

\author{Jeffrey Neethi Neethirajan}
\email{Jeffrey.Neethirajan@cpfs.mpg.de}
\affiliation{
Max Planck Institute for Chemical Physics of Solids, 01187 Dresden, Germany
}

\author{Benedikt Daurer}%
\affiliation{%
Diamond Light Source, Harwell Science and Innovation Campus, Didcot, 0X11 ODE, United Kingdom
}%

\author{Marisel Di Pietro Mart\'inez}
\affiliation{
Max Planck Institute for Chemical Physics of Solids, 01187 Dresden, Germany
}

\author{Ale$\check{\textrm{s}}$ Hrabec}
\affiliation{
Laboratory for Mesoscopic Systems, Department of Materials, ETH Zurich, 8093 Zurich, Switzerland
}%
\affiliation{
Laboratory for Multiscale Materials Experiments, Paul Scherrer Institute, 5232 Villigen PSI, Switzerland
}%

\author{Luke Turnbull}
\affiliation{
Max Planck Institute for Chemical Physics of Solids, 01187 Dresden, Germany
}

\author{Majid Kazemian}
\affiliation{%
Diamond Light Source, Harwell Science and Innovation Campus, Didcot, 0X11 ODE, United Kingdom
}%

\author{Burkhard Kaulich}
\affiliation{%
Diamond Light Source, Harwell Science and Innovation Campus, Didcot, 0X11 ODE, United Kingdom
}%

\author{Claire Donnelly}
\email{Claire.Donnelly@cpfs.mpg.de}
\affiliation{
Max Planck Institute for Chemical Physics of Solids, 01187 Dresden, Germany
}%
\affiliation{International Institute for Sustainability with Knotted Chiral Meta Matter (WPI-SKCM$^2$), Hiroshima University, Hiroshima 739-8526, Japan}
\noaffiliation

\date{\today}

\begin{abstract}
Imaging of nanoscale magnetic textures within extended material systems is of critical importance both to fundamental research and technological applications. Whilst high resolution magnetic imaging of thin nanoscale samples is well-established with electron and soft X-ray microscopy, the extension to micrometer-thick systems with hard X-rays currently limits high resolution imaging to rare-earth magnets. Here we overcome this limitation by establishing soft X-ray magnetic imaging of micrometer-thick systems using the pre-edge phase X-ray Magnetic Circular Dichroism signal, thus making possible the study of a wide range of magnetic materials. By performing dichroic spectro-ptychography, we demonstrate high spatial resolution imaging of magnetic samples up to 1.7\,$\mu$m thick, an order of magnitude higher than conventionally possible with absorption-based techniques. This new regime of magnetic imaging makes possible the study of extended non rare-earth systems that have until now been inaccessible, from magnetic textures for future spintronic applications to non-rare-earth permanent magnets. 
\end{abstract}

\maketitle

Magnetic materials have a high impact on our society, with a range of functionalities making possible a number of applications. On one hand, the study of naturally occurring magnetite in the natural world gives insight into magnetoreception \cite{magnetite_magnetoreception} and the role of the earth’s field over the ages \cite{earth_magnetic_field_rich}. On the other hand, strong magnetic fields from highly anisotropic permanent magnets play a key role in the production of clean energy \cite{perm_mag, mems_energy, perm_magnet_1},  highly inductive magnets play an important role in write heads in hard disk drives \cite{write_head_inductive_magnet} and topological textures in spintronics devices promise the next generation of computing technologies \cite{parkin_racetrack_2008,fert_skyrmion_racetrack}. 

Key to the behaviour of these magnetic systems is their underlying magnetisation configuration, which forms local areas of uniform magnetisation – called magnetic domains – as well as nanoscale topological magnetisation defects such as domain walls. Direct imaging of the magnetisation configuration provides a unique insight into the underlying mechanisms responsible for these behaviours. For example, imaging the reversal processes of permanent magnets elucidates their switching mechanisms \cite{3d_takeuchi_permanentmagnets}, allowing for the development of more efficient devices, while imaging of the propagation of topological magnetisation textures has allowed for the step towards realising devices based on the ultra-fast motion of magnetic defects through interconnected networks  \cite{blowing_skyrmion_bubbles, parkin_racetrack_2008} and non-linear dynamics enabling new types of computing architectures \cite{skyrmion_neuromorphic}.

With such a diverse variety of magnetic systems, we require a broad range of capabilities to image the underlying magnetic configurations. First, we require the ability to study systems of varying dimensions, from single atoms to thick magnetic systems. Second, we require flexibility to study a wide variety of materials, from naturally forming magnetite, to exotic designed topological chiral magnets. And last, we require sufficient spatial resolution to resolve magnetic textures on the order of the magnetic exchange length: down to tens of nanometres and below, which corresponds to the typical sizes of key topological textures such as domain walls, skyrmions and even hopfions.

However, while for thin samples ($<200\,$nm) and surfaces, material flexibility and spatial resolution are well established with high spatial resolution soft X-ray \cite{soft_xray_imaging_10nm_PFischer, aurelio_soft_Xray_3d} and electron microscopies \cite{LTEM_review, skyrmion_tube}, the imaging of thicker extended systems is more challenging. High spatial resolution imaging of extended samples of thicknesses on the order of hundreds of nanometres to micrometres has been achieved with resonant hard X-ray dichroic imaging \cite{claire_prb} which, when combined with tomographic imaging, has revealed singularities known as Bloch points, skyrmions and magnetic vortex rings \cite{claire_bp_first, claire_vortex_ring, claire_time_resolved_3d, seki_scalar_XMCD_tomo} within micrometre-thick samples with spatial resolutions down to 50\,nm \cite{claire_time_resolved_3d}. However, in the hard X-ray regime, X-ray dichroic signals are highly material dependent: while relatively strong signals exist for certain materials such as rare-earth containing compounds,  hard X-ray dichroic signals of transition metal magnets are approximately $20\times$ weaker \cite{claire_prb}. This significantly weaker dichroic signal results in  poorer spatial resolution and imaging quality, thus generally limiting 3D investigations in these materials to thin films.
 
Here we establish a route to the high spatial resolution magnetic imaging of extended magnetic systems that is applicable to a wide range of magnetic materials with soft X-rays. By exploiting the phase XMCD signal, which is prominent in the pre-absorption edge, we extend soft X-ray magnetic imaging at transition metal edges to samples an order of magnitude thicker than currently viable, opening up a new regime for the imaging of magnetic compounds.
We gain access to the phase contrast using X-ray ptychography, a coherent diffractive imaging (CDI) technique. By performing dichroic spectro-ptychography we map out the complex XMCD signal, revealing a notable phase XMCD signal in the pre-absorption edge where the absorption contrast vanishes. This pre-edge phase signal makes it possible to measure thicker samples, and in this way we image the magnetic configuration of samples up to 1.7$\,\mu$m in thickness: an order of magnitude higher than what is typically measured with soft X-rays.

The limitation of soft X-ray magnetic imaging to thin samples exists due to magnetic scattering being a resonant effect. Indeed, when a photon energy is tuned close to an absorption edge between a core level and a magnetically polarised valence band, the electronic transition between the two bands is dependent on the helicity of the incoming photon and the projection of the magnetization vector along the direction of propagation of X-rays \cite{xmcd_1,xmcd_2,xmcd_3}. Experimentally, this leads to differences in absorption of the X-rays which, when combined with nano-microscopy, can provide projections of the magnetization in a sample. When compared to electron microscopy, which is limited to thin samples on the order of hundred nanometres in thickness, soft X-rays provide a higher penetration depth while being element specific. However, for magnetic imaging, the localisation of measurable XMCD signals to resonance energies where the high absorption can lead to zero transmission for extended thick systems, have meant that both soft X-rays and electron imaging have generally been limited to thin films on the order of hundreds of nanometers. 
\begin{figure}[htbp]
  \centering
  \includegraphics[width=.5\textwidth, keepaspectratio=True]{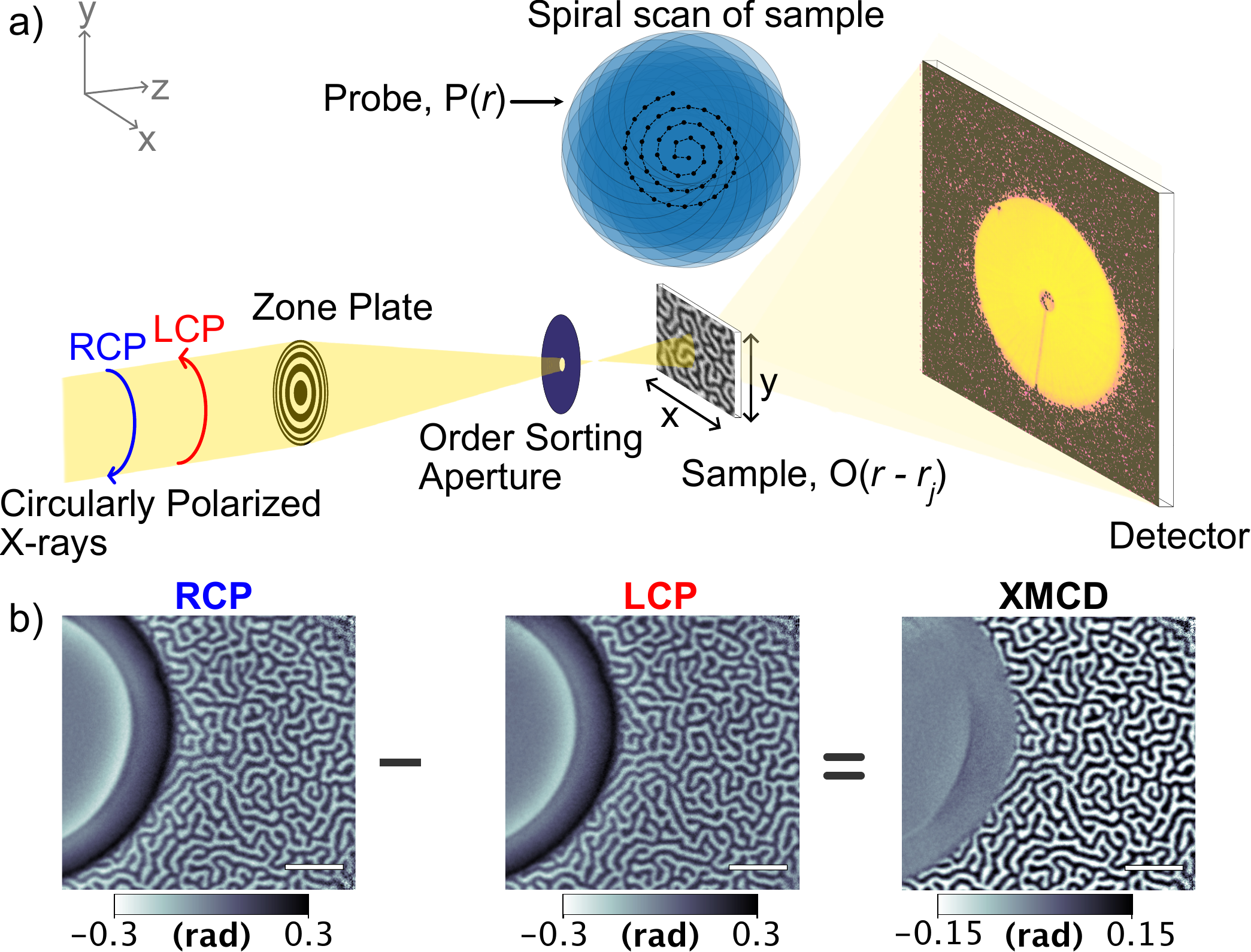}
  \caption{a) Schematic of the ptychography setup. The sample is scanned in the plane perpendicular to the direction of propagation of the X-rays, for overlapping probe illumination positions, as indicated in the inset. Diffraction patterns are measured in the far field. b) Ptychographic phase projections measured with RCP and LCP X-rays, with the difference giving the XMCD signal, indicating the projection of the magnetisation (anti)parallel to the X-ray beam. The semicircle is a hole milled using a focused Ga ion beam which provides an empty region of the sample for image normalisation and alignment. Scale bar represents 1\,$\mu$m.}
  \label{setup}
\end{figure}

\begin{figure*}[htbp]
  \centering
  \includegraphics[width=\textwidth, keepaspectratio=True]{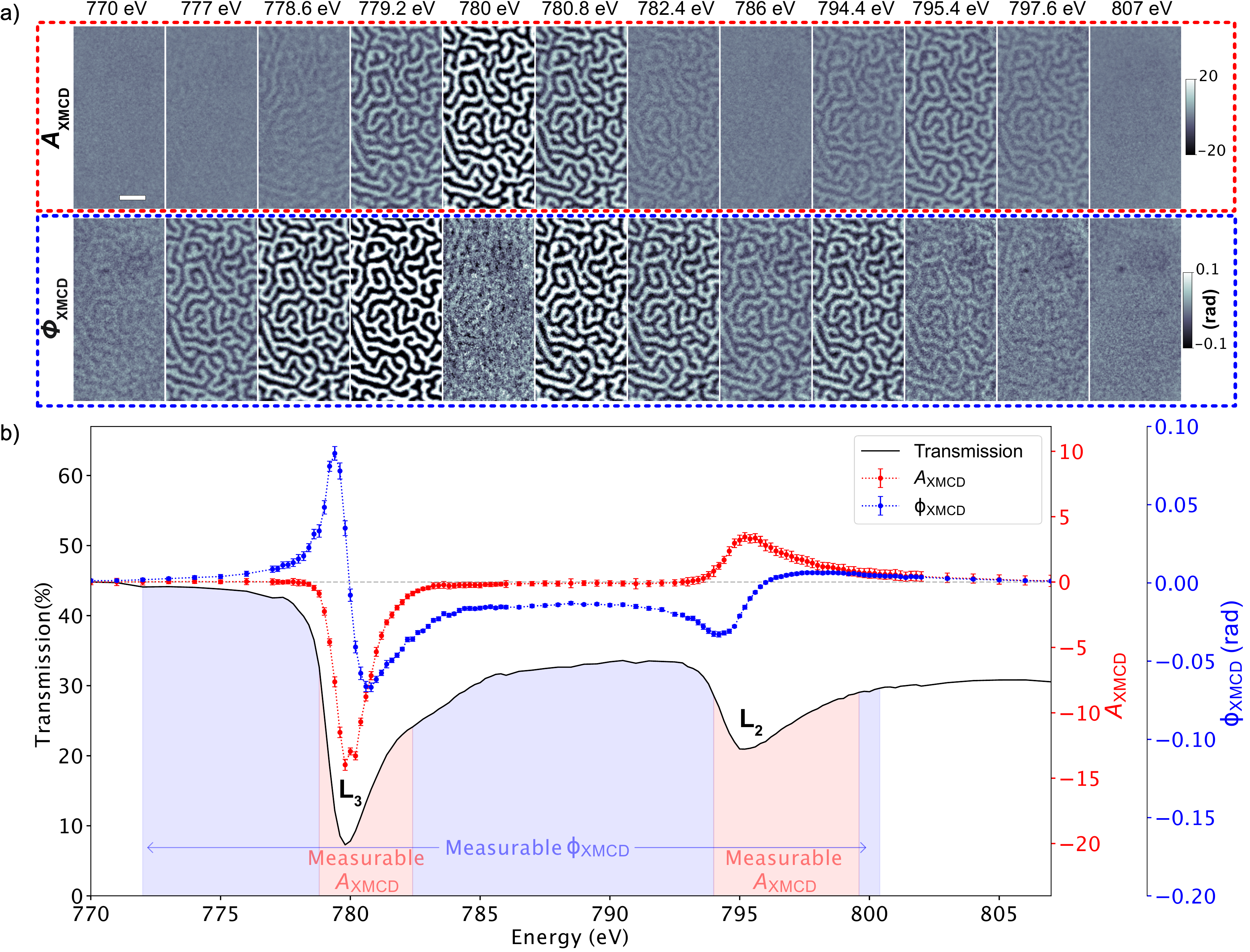}
  \caption{Dichroic spectro-ptychography across the Co L$_3$ and L$_2$ absorption edges of 100\,nm thick CoPt multi-layer system. a) The amplitude ($A_{XMCD}$) and phase ($\phi_\mathrm{{XMCD}}$) XMCD images across a selected range of energies. By definition, $A_\mathrm{{XMCD}}$ is dimensionless and $\phi_\mathrm{{XMCD}}$ is in radians. Scale bar represents 500\,nm. b) $A_{\mathrm{XMCD}}$ (red) and $\phi_{\mathrm{XMCD}}$ (blue) XMCD spectra across the Co L$_{2,3}$ edges extracted from domains seen in the XMCD images plotted in a). The black curve represents the transmission through the thickness of the sample. The $\phi_{\mathrm{XMCD}}$ spectra exhibits a non-zero signal throughout the entire range of energy except on resonance. The $A_{\mathrm{XMCD}}$ signal is notably strong only on resonance, when transmission is minimum and vanishes at energies 1.8 eV off-resonance. The blue shaded region represents areas where only $\phi_{\mathrm{XMCD}}$ is measurable and red shaded regions represent areas where $A_{\mathrm{XMCD}}$ is also measurable.}
  \label{spectro_ptycho}
\end{figure*}

However, the magnetic contrast does not only present itself in the absorption: the scattering factor, and therefore the refractive index of a material, is complex, meaning that magnetic dichroism is also present in the phase of the transmitted wave, see Appendix \ref{complex_xmcd} for more details. With the development of lensless CDI techniques such as holography and ptychography, the full complex transmission function of an object becomes experimentally accessible with the help of phase retrieval algorithms \cite{fienup_phase_algo}. These lensless imaging techniques have revolutionised X-ray imaging, with phase imaging of weakly absorbing objects \cite{phase_imaging_1, phase_imaging_2} opening up the possibility to image biological samples \cite{phase_imaging_holo}, and the prospect of diffraction-limited spatial resolutions \cite{smco5, magneto_bacteria}. 
So far magnetic imaging has mainly benefited from the high spatial resolutions that are available with lensless CDI techniques, bringing spatial resolutions down to 10\,nm \cite{smco5} and below \cite{magneto_bacteria}. However, although it has been seen that the phase dichroism exists across the energy spectrum \cite{claire_prb, phase_imaging_holo}, offering possibilities for low radiation dose imaging \cite{phase_imaging_holo}, so far magnetic phase imaging has not yet been fully exploited. \par
This X-ray phase dichroism is exploited here to extend the applicability of soft X-ray magnetic imaging to thicker systems. We first map the complex XMCD signal by imaging 100\,nm thick CoPt multilayer system with perpendicular anisotropy. The multilayer film was grown by magnetron sputtering on an X-ray transparent silicon nitride membrane (see Appendix \ref{sample_fab}). Subsequently, a hole of diameter $\sim 3\,\mu$m in size milled using a focused Ga ion beam in order to provide an empty region of the sample for image normalisation and alignment. Dichroic spectro-ptychography was performed at the i08-1 beamline of the Diamond Light Source to map the complex XMCD signal from the magnetic domains in the sample. Specifically, circularly polarised X-rays were microfocused on to the Sample (O(\textit{r})) as it was scanned for several overlapping Probe (P\textit{r}) positions, $r_j$, in the plane perpendicular to the direction of propagation of the X-rays. For each position, a coherent diffraction pattern is collected in the far field by a two dimensional detector. The complex transmission function of the object is then retrieved iteratively with the help of reconstruction algorithms \cite{ptycho_ref, ptypy}.  A simple schematic of the experimental setup is shown in Fig.\,\ref{setup}a and further experimental details are explained in Appendix\,\ref{setup and reconstructions}.

\begin{figure*}[htbp]
  \centering
  \includegraphics[width=\textwidth, keepaspectratio=True]{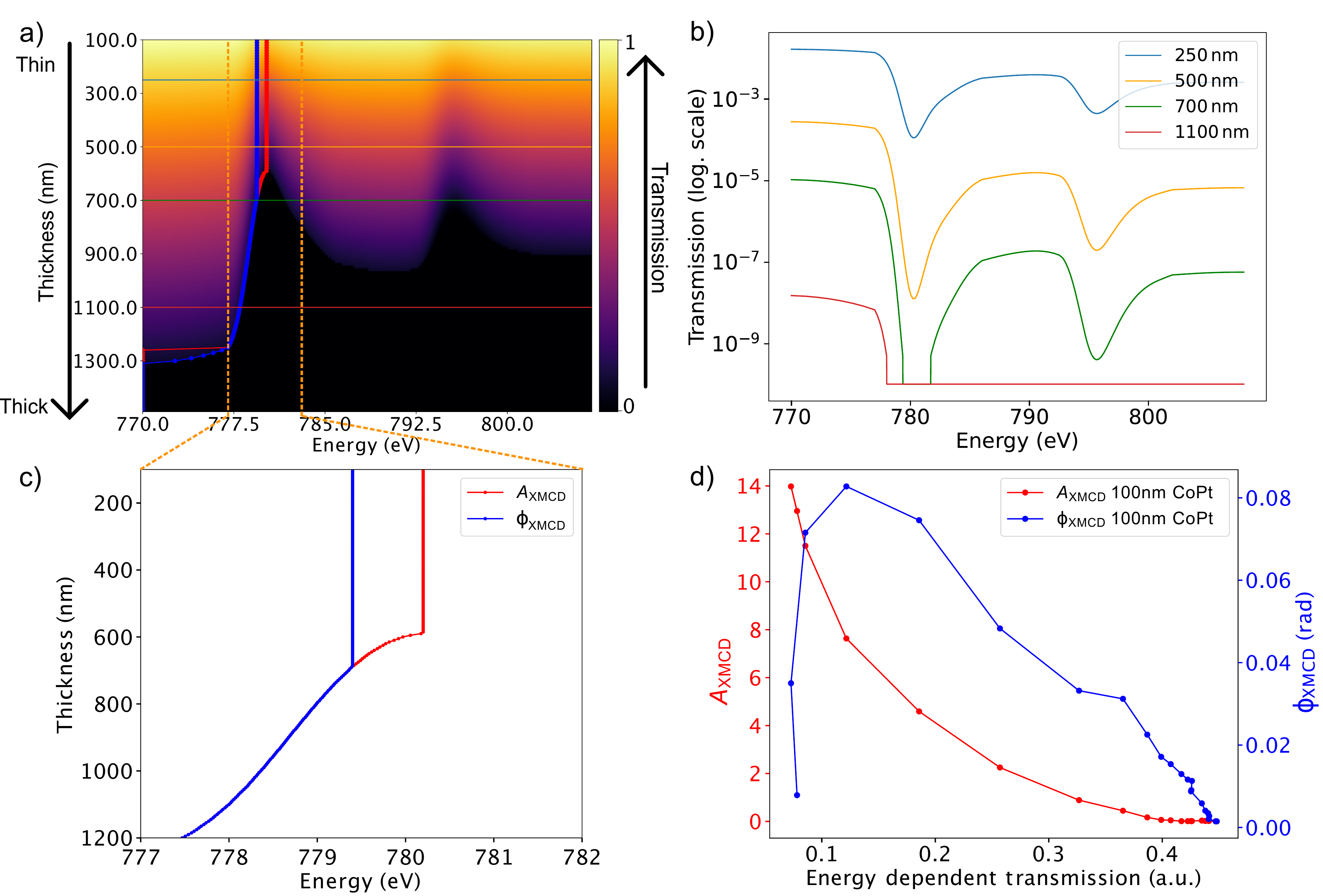}
  \caption{a) Calculated transmission of thicker CoPt films as a function of energy across the Co L$_3$ and L$_2$ absorption edges. The darker regions indicate regions of low or zero transmission. As the thickness is increased we are forced to measure at pre-absorption edge energies where transmission is significantly higher than on-resonance, indicated by the red and blue dotted lines for $A_{\mathrm{XMCD}}$ and $\phi_{\mathrm{XMCD}}$ respectively. b) Transmission spectra for a few select calculated thicknesses taken across the line corresponding to the lines shown in (b). c) Optimal energies at which highest $A_{\mathrm{XMCD}}$ and $\phi_{\mathrm{XMCD}}$ can be extracted as sample thickness is increased. d) Evolution of the complex pre-edge XMCD signal as a function of energy-dependent transmission of the material. As one moves to energies corresponding to higher transmission, $A_{\mathrm{XMCD}}$ exhibits an exponential decay while the $\phi_{\mathrm{XMCD}}$ exhibits a linear decay.}
  \label{signal_vs_thickness}
\end{figure*}

The reconstructed phase images taken using Right Circular Polarised (RCP) and Left Circular Polarised (LCP) X-rays are shown in Fig.\,\ref{setup}b, and reveal a labyrinth-like domain structure in the CoPt film, with the magnetisation in the domains oriented perpendicular to the sample plane, and (anti)parallel to the direction of the propagation of the X-rays. When the polarisation is changed from RCP to LCP, the XMCD contrast switches, allowing for the isolation of the magnetic signal as shown in Figure\,\ref{setup}b. We define the amplitude XMCD ($A\mathrm{_{XMCD}}$) and phase XMCD ($\mathrm{\phi_{XMCD}}$) signal as follows: \begin{equation}
     A_{\mathrm{XMCD}} = \frac{\log_{e} (A_{\mathrm{RCP}})  - \log_{e} (A_\mathrm{{LCP}})}{2}
    \label{xmcd_def}
 \end{equation}
 \begin{equation*}
      \mathrm{\phi_{XMCD} = \frac{\phi_{RCP} - \phi_{LCP}}{2}}
 \end{equation*}
where $A\mathrm{_{RCP}}$, $A\mathrm{_{LCP}}$, $\mathrm{\phi_{RCP}}$ and $\mathrm{\phi_{LCP}}$ are the amplitude and phase projections taken with RCP and LCP X-rays respectively.
In order to map the complex XMCD signal across the L$_3$ and L$_2$ edges, dichroic spectro-ptychography scans were performed for a range of energies between 770\,eV and 807\,eV in the vicinity of the Co L$_3$ and L$_2$ absorption edges with RCP and LCP X-rays. The dichroic $A\mathrm{_{XMCD}}$ and $\mathrm{\phi_{XMCD}}$ images for a select few energies are given in Fig.\,\ref{spectro_ptycho}a. To extract the complex XMCD signal, the domains are segmented and their contrast averaged (details explained in Appendix \ref{image_analysis}) to obtain the quantitative spectra plotted in Fig.\,\ref{spectro_ptycho}b. The solid black curve, shown in Fig.\,\ref{spectro_ptycho}b, represents the transmission spectrum through the sample, providing a direct comparison between the strength of the respective XMCD signals and transmission through the sample. We first consider the $A_{\mathrm{XMCD}}$, with the images highlighted by the red dotted box in the top row of Fig.\,\ref{spectro_ptycho}a, and the extracted $A_{\mathrm{XMCD}}$ spectrum given by the red curve in Fig.\,\ref{spectro_ptycho}b. Two resonance peaks of opposite sign can be observed across the L$_3$ and L$_2$ edges, where domain contrast reversal can also be seen in the $A_{\mathrm{XMCD}}$ images. The signal and the domain contrast is strongest at the energy associated with highest absorption (780\,eV), and  already drops to zero, 1.8\,eV away from resonance with the magnetic domains no longer resolvable in the images.

We next consider the $\mathrm{\phi_{XMCD}}$ signal, represented by the blue curve in Fig.\,\ref{spectro_ptycho}b. Corresponding domain contrast can be observed in almost all of the $\mathrm{\phi_{XMCD}}$ projections at different energies, highlighted in a blue dotted box in Fig\,\ref{spectro_ptycho}a. The $\mathrm{\phi_{XMCD}}$ signal is particularly strong in the vicinity of the L$_3$ and L$_2$ edges, with the maximum occurring 0.5\,eV below the absorption edge, with images showing strong contrast around 779.4\,eV and 794.4\,eV. In the pre-absorption edge, the $\mathrm{\phi_{XMCD}}$ domain contrast is opposite to $A_{\mathrm{XMCD}}$, while we observe a $\mathrm{\phi_{XMCD}}$ contrast reversal across the two absorption edges. Most notably, the difference between the $\mathrm{\phi_{XMCD}}$ and $A_{\mathrm{XMCD}}$ is that while the $A_{\mathrm{XMCD}}$ is restricted to on-resonance energies, the $\mathrm{\phi_{XMCD}}$ signal is non-zero across almost all energies measured, specifically in the pre-absorption edge where the magnetic domains can even be resolved 10\,eV below the absorption edge, and transmission through the sample is significantly higher. To visualise this difference in the energy-dependent measurable contrast, we have shaded the regions of the spectrum in Fig.\,\ref{spectro_ptycho}b, where red regions indicate the energies for which the $A_{\mathrm{XMCD}}$ is measurable (as well as the $\mathrm{\phi_{XMCD}}$), while blue regions indicate the energy regime for which only the $\mathrm{\phi_{XMCD}}$ is detectable. The $\mathrm{\phi_{XMCD}}$ is available for a much wider range of available energies than the $A_{\mathrm{XMCD}}$, offering a more flexible contrast mechanism.

The importance of the flexibility of the contrast mechanism becomes clear when we consider thicker systems. While it is clear that on-resonance imaging with $A_{\mathrm{XMCD}}$ contrast works well for thin samples, as we increase the thickness above a threshold at which there is not sufficient transmission on resonance for measurements, we will be forced to image off-resonance at energies associated with lower absorption. We explore this by calculating the transmission for thicker samples using the experimentally measured transmission spectrum of the CoPt multilayer and further calculating the complex XMCD signal, as explained in Appendix \ref{simulation_thick_systems}. We first consider the calculated transmission for thicker samples, shown in Fig.\,\ref{signal_vs_thickness}a, with each row in the image representing transmission as a function of energy for a sample of a certain thickness. The darker regions indicate a lower transmission through the sample and we observe a reduction in energies with significant transmission as the thickness increases. Examples of transmission spectra corresponding to selected thicknesses (indicated as coloured lines on Fig.\,\ref{signal_vs_thickness}a) are given in Fig.\,\ref{signal_vs_thickness}b where one can observe a drop to zero transmission on-resonance for higher thicknesses.
\begin{figure*}[htbp]
  \centering
  \includegraphics[width=.8\textwidth, keepaspectratio=True]{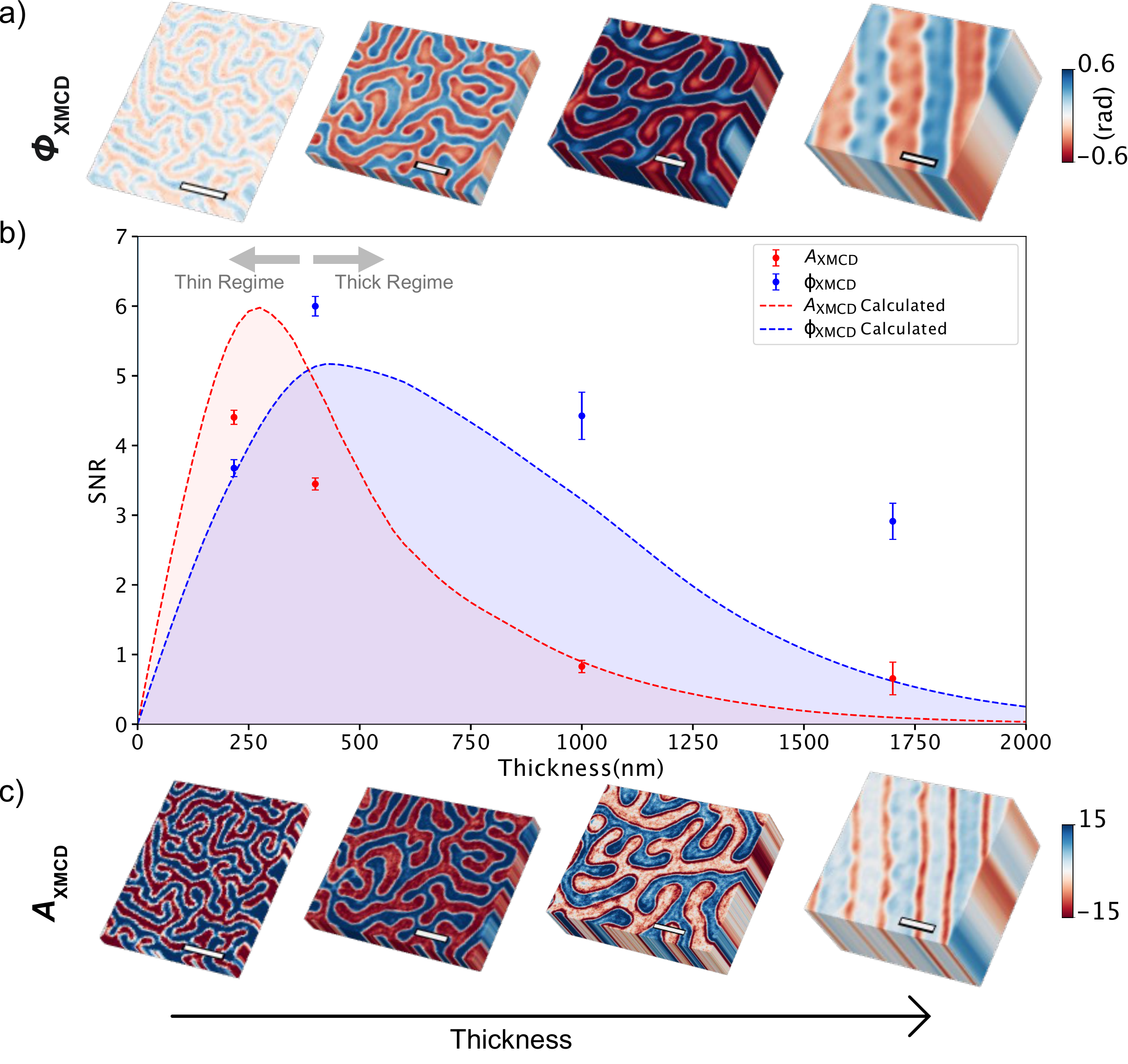}
  \caption{a) and c) $\phi_{\mathrm{XMCD}}$ and  $A_{\mathrm{XMCD}}$ imaging of magnetic films of increasing thickness. The XMCD projections with highest SNR of the 100\,nm thick CoPt  were measured at 780\,eV and 779.4\,eV for $A_{\mathrm{XMCD}}$ and $\phi_{\mathrm{XMCD}}$ respectively. The XMCD projections with highest SNR for the 400\,nm, 1\,$\mu$m and 1.7\,$\mu$m FeGd films were measured at 709\,eV, 708.5\,eV and 708\,eV, respectively, for both $A_{\mathrm{XMCD}}$ and $\phi_{\mathrm{XMCD}}$. b) Calculated highest SNR of $A_{\mathrm{XMCD}}$ and $\phi_{\mathrm{XMCD}}$ for a range of thicknesses indicated by the red and blue dashed line, respectively. Note that the calculated thickness dependence of the CoPt XMCD SNR is scaled in thickness to allow for a comparison between the experimentally measured CoPt/FeGd data shown as red and blue dots (see Appendix \ref{simulation_thick_systems}). We identify a 'thin' regime where the $A_{\mathrm{XMCD}}$ SNR provides a significantly better image quality, and a 'thick' regime, where due to the pre-edge $\phi_{\mathrm{XMCD}}$ we obtain high quality projections for thick samples with high SNR. The highest $A_{\mathrm{XMCD}}$ and $\phi_{\mathrm{XMCD}}$ SNR across the Co L$_{2,3}$ edges for 100\,nm thick CoPt (scaled to an effective thickness of 217\,nm FeGd), and 400\,nm , 1\,$\mu$m and 1.7\,$\mu$m FeGd. The extracted optimal SNR values from each of these XMCD projections is plotted, with red circular dots indicating the  $A_{\mathrm{XMCD}}$ SNR and  blue squares indicating the $\phi_{\mathrm{XMCD}}$ SNR for the different thicknesses. The images shown in a) and c) are the projections from which the $A_{\mathrm{XMCD}}$ and $\phi_{\mathrm{XMCD}}$ signal were extracted. Scale bar in the images represents 500\,nm.}
  \label{thick_imaging}
\end{figure*}

We next consider the complex XMCD signal for thicker samples, by plotting the energy at which both strongest signals can be extracted for increasing thickness in Fig.\,\ref{signal_vs_thickness}c. We observe that above a threshold thickness, the optimal energy at which the XMCD can be measured drops steadily. This decrease in the measurement energy, to energies associated with higher transmission, has further implications when we consider the dependence of the  $A_{\mathrm{XMCD}}$ and $\mathrm{\phi_{XMCD}}$ signal on the energy-dependent transmission. Indeed, by plotting the complex pre-edge XMCD signal, measured for 100\,nm CoPt shown in Fig.\,\ref{spectro_ptycho}b, as a function of transmission in Fig.\,\ref{signal_vs_thickness}d, we observe a significant difference: the $A_{\mathrm{XMCD}}$ is maximum for lower transmission, and decays exponentially as the transmission is increased. In contrast, the $\mathrm{\phi_{XMCD}}$ signal is maximum for energies corresponding to higher transmission, and decays linearly as the transmission is increased, thus exhibiting a weaker dependence on the transmission of the sample. For purely absorption-based imaging, the requirement to measure at lower off-resonance energies thus quickly leads to the suppression of the XMCD signal, however access to the pre-edge $\mathrm{\phi_{XMCD}}$ opens the possibility to measure thicker samples.

To explore the evolution of the $\mathrm{A_{XMCD}}$ and $\mathrm{\phi_{XMCD}}$ for increasing thickness, we investigate the dimensionless Signal to Noise ratio (SNR) for the calculated $A_{\mathrm{XMCD}}$ and $\mathrm{\phi_{XMCD}}$ as a function of effective thickness (scaled for comparison with experimental data, see Appendix \ref{simulation_thick_systems}), plotted in Fig.\,\ref{thick_imaging}b (dashed lines). For thinner samples, where the $A_{\mathrm{XMCD}}$ SNR is significantly higher than the $\mathrm{\phi_{XMCD}}$ SNR, the $A_{\mathrm{XMCD}}$ SNR peaks at a thickness corresponding to the suppression of on-resonance transmission, and decreases exponentially afterwards. However, the $\mathrm{\phi_{XMCD}}$ SNR increases at a slower rate, peaking at a higher effective thickness, but retaining a high SNR to yet higher thicknesses. We identify a `thin' regime where the $A_{\mathrm{XMCD}}$ SNR provides a significantly better image quality, and a `thick' regime, where $\phi_{\mathrm{XMCD}}$ SNR provides high quality images.

We experimentally confirm this ability to image the magnetic state of thicker samples with high SNR $\phi_{\mathrm{XMCD}}$ imaging by performing dichroic spectro-ptychography scans on FeGd films grown by magnetron co-sputtering of thicknesses 400\,nm, 1\,$\mu$m and 1.7\,$\mu$m across the the L$_3$ and L$_2$ edges of Fe (refer Appendix \ref{sample_fab} and \ref{FeGd_1000nm_spectro_ptycho} for more details). We first plot the highest SNR images for $\mathrm{\phi_{XMCD}}$ and  $A_{\mathrm{XMCD}}$ and in Fig.\,\ref{thick_imaging}a and \ref{thick_imaging}c respectively, where it can be seen that the quality of the images is highly thickness dependent and that this thickness dependence is different for the two contrast mechanisms. In particular, for the $A_{\mathrm{XMCD}}$ the quality of the image appears to decrease steadily with thickness, with a loss of quantitative XMCD signal for thicknesses of 1\,$\mu$m and above, with contrast only present in the vicinity of the domain walls. For the $\mathrm{\phi_{XMCD}}$, the maximum quality instead appears to occur at 1$\,\mu$m, and the quantitative measurement of the magnetic domain structure (indicated by the equal and opposite magnetic contrast in positive and negative domains) is retained for all sample thicknesses. Furthermore, the $\mathrm{\phi_{XMCD}}$ images have a higher spatial resolution than the $A_{\mathrm{XMCD}}$ images for samples in the `thick regime', see Appendix \ref{FRC_spatial_res} for more details.

To quantitatively compare both $A_{\mathrm{XMCD}}$ and $\mathrm{\phi_{XMCD}}$ for the various thicknesses in these images, we calculate the dimensionless Signal to Noise Ratio (SNR) of the measured $A_{\mathrm{XMCD}}$ and $\mathrm{\phi_{XMCD}}$ projections, combining our 100\,nm CoPt and thicker FeGd films by defining an effective thickness as discussed in the Appendix \ref{simulation_thick_systems}. The SNR of the images is plotted as a function of effective thickness in Fig\,\ref{thick_imaging}b, with red and blue dots representing the measured $A_{\mathrm{XMCD}}$ and $\mathrm{\phi_{XMCD}}$ SNR. As observed in the images, we see a clear difference in the thickness dependence, with the $\mathrm{\phi_{XMCD}}$ providing high SNR imaging of magnetic domains for thicknesses up to 1.7\,$\mu$m, the thickest film that was measured. 

By comparing with our calculated SNR plotted in Fig.\,\ref{thick_imaging}b (dashed lines), it can be seen that the $\mathrm{\phi_{XMCD}}$ imaging can extend to samples up to 2$\,\mu$m, an order of magnitude thicker than what is currently achievable with soft X-ray absorption imaging. While this exact thickness dependence is highly material dependent, this new approach addresses a key limitation of current imaging capabilities, making possible the high spatial resolution mapping of thicker transition metal-based systems that until now have been inaccessible. 

In conclusion, we have demonstrated the soft X-ray magnetic imaging of thick magnetic systems by imaging in the pre-edge $\mathrm{\phi_{XMCD}}$ signal. By determining the complex XMCD spectrum of a CoPt thin film across the L$_3$ and L$_2$ absorption edges of Co, we were able to not only identify the presence of a significant $\phi_{\mathrm{XMCD}}$ signal in the pre-edge regime but, by extrapolating this data, also to establish a new imaging regime where the $\phi_{\mathrm{XMCD}}$ signal enables quantitatively imaging of thick samples with high SNR and spatial resolution that would not be possible with traditional absorption based techniques. Remarkably, our analysis predicts an order of magnitude increase in the accessible thickness regime due to the phase imaging, which we confirmed by imaging magnetic domains in FeGd samples of up to 1.7$\,\mu$m  in thickness. 

We expect pre-edge magnetic phase imaging to have a significant impact on the field of magnetism, making possible the imaging of topological defects in higher dimensional chiral magnets, where this new-found flexibility in the material and sample geometry will drive forward the discovery of exotic textures. Moreover, it will enable the non-destructive investigation of naturally occurring magnetic systems, providing insight into the formation and role of magneto-fossils \cite{magnetofossils_2008} and meteorites \cite{meteorites_magnetic_imaging}. Finally, an immediate societal impact will be found with the study of materials critical to efficient and clean energy production, opening the door to the mapping of the internal configuration of non-rare earth magnets \cite{non_rare_earth_perm_mag_future}. 

\section*{Acknowledgements}
Diamond Light Source provided access to the i08-1 Soft X-ray Ptychography Facility with experiment grants MG32635-1 and MG28255-1. J.N.N., M.D.P.M, L.T. and C.D. acknowledge funding from the Max Planck Society Lise Meitner Excellence Program. J.N.N acknowledges support from the International Max Planck Research School for Chemistry and Physics of Quantum Materials.

%\bibliography{main_bbl}% Produces the bibliography via BibTeX.
%apsrev4-2.bst 2019-01-14 (MD) hand-edited version of apsrev4-1.bst
%Control: key (0)
%Control: author (8) initials jnrlst
%Control: editor formatted (1) identically to author
%Control: production of article title (0) allowed
%Control: page (0) single
%Control: year (1) truncated
%Control: production of eprint (0) enabled
\providecommand{\noopsort}[1]{}\providecommand{\singleletter}[1]{#1}%

\appendix

\section{Complex X-ray magnetic circular dichroism}
\label{complex_xmcd}
The complex reconstructed images obtained by ptychography can be expressed as $A(E) e^{i\phi(E)}$ where both Amplitude ($A$) and Phase ($\phi$) are both highly energy dependent and embed information about the complex scattering factor given by,
\begin{equation*}
     f(E, r) = \underbrace{f_c(E)\cdot(\varepsilon ^*_f \cdot \varepsilon _i)}_\text{Electronic} - \underbrace{i f_1(E) \cdot (\varepsilon ^*_f \times \varepsilon _i) \textbf{m(r)}}_\text{Circular Dichroism}
\end{equation*}
where E is the energy, $r$ is the position vector, $\varepsilon_f$ and $\varepsilon_i$ are the initial and final Jones polarisation vectors, $f_c(E)$ is the electronic scattering factor, $f_1(E)$ is the magnetic scattering factor and $\textbf{m(r)}$ is the orientation of magnetisation. For the case of circularly polarised light, the complex scattering factor reduces to:
\begin{equation*}
     f(E, r) = \underbrace{f_c(E)}_\text{Electronic} \pm \underbrace{i f_1(E)  \textbf{m(r)}\cdot \mathbf{\hat{k}}}_\text{Circular Dichroism}
\end{equation*}
where $\mathbf{\hat{k}}$ is the wave number of X-rays. 
Further details on the relationship between $A$ and $\phi$ to the scattering factor is explained in reference \cite{claire_prb}. Given that $A(E) \propto e^{\Im(f_1(E)) \textbf{m} \cdot \mathbf{\hat{k}}}$, the logarithmic difference between RCP and LCP gives the pure magnetic scattering factor and we define our $A_{\mathrm{XMCD}}$ signal as:
\begin{align*}
    A_{\mathrm{XMCD}} =& \frac{\log_{e}(A_\mathrm{{RCP}}) - \log_{e}(A_\mathrm{{LCP}})}{2} \\
    &\propto \,\Im (f_1(E)) \mathbf{m} \cdot \mathbf{\hat{k}}
\end{align*}
while the direct difference between the RCP and LCP phase images gives the $\phi_{\mathrm{XMCD}}$ signal:
\begin{align*}
    \phi_{\mathrm{XMCD}} &= \phi_{\mathrm{RCP}}-\phi_{\mathrm{LCP}}\\
    &\propto \,\Re (f_1(E))\textbf{m} \cdot \mathbf{\hat{k}}
\end{align*}

\section{Sample Fabrication}
\label{sample_fab}
The 100\,nm CoPt film was grown with the following composition $\mathrm{Ta\backslash Pt(4nm)\backslash {Co(1nm)\backslash Pt(1nm)}\times50\backslash Ru(3nm)}$  by magnetron sputtering on an X-ray transparent $\mathrm{5\,mm\times5\,mm}$ silicon nitride ($\mathrm{Si_3N_4}$) membrane
The magnetic Fe:Gd (70:30 at\%) films of various thicknesses were deposited via magnetron co-sputtering on  $\mathrm{5\,mm\times5\,mm}$ $\mathrm{Si_3N_4}$ windows at base pressure of $1\times10^{-8}$ Torr and Ar sputtering atmosphere pressure of 3\,mTorr using commercial sputtering system.
Holes in the films were milled with a focused Ga ion beam to provide a region for alignment and normalisation.

\section{Ptychography setup and reconstruction}
\label{setup and reconstructions}
For each ptychography scan, diffraction patterns were recorded on a 2048\,x\,2048 pixel sCMOS area detector (AXIS-SXRF-2020EUV, Axis Photonique Inc.) with an exposure of 40\,ms while laterally scanning the sample with a Piezo scanner in a spiral pattern across the X-ray beam, to minimise grid pathology artefacts. The sCMOS camera with a pixel size of 6.5\,$\mu$m was placed approximately 72\,mm downstream of the sample. The sample was placed about 70$\,\mu$m downstream of the focus formed by a Fresnel Zone Plate (FZP) with a diameter of 333$\,\mu$m and a focal length of 13.725\,mm, producing a defocused beam with a full-width-at-half-maximum (FWHM) of approximately 1\,$\mu$m. Together with a scanning step size of 200\,nm, this gives an overlapping ratio of about 80\,\%. At the beginning of each scan, a single dark image was collected which was  subtracted from all raw diffraction images of that scan. In addition, all raw images were loss-lessly reduced to a 512x512 image by first cropping the central 1024x1024 pixels and then binning the image by a factor of 2 along each dimension. The ptychographic reconstructions were performed using the PtyPy software \cite{ptypy}, loading the clean dark-subtracted 512x512 diffraction images together with their corresponding scan positions as recorded by an interferometric system and subsequently running 1000 iterations of the Graphical Processing Unit -accelerated implementation of the Difference Map algorithm. All images were reconstructed with a fixed pixel size of 17.414\,nm, to account for the energy-dependent pixel size of the dichroic spectro-ptychography scans. This value corresponds to the pixelsize of the ptychographic reconstruction associated with the lowest photon energy.

 \section{Image Analysis}\label{image_analysis}
 The images undergo a preprocessing routine to quantitatively extract the magnetic contrast. All images are aligned with respect to a hole on the sample (shown in Fig\,\ref{setup}b) and all phase projections were corrected for a linear phase ramp. Additionally, to filter low frequency noise from the phase images, a gaussian filter with high sigma is applied to blur the phase image, which is then subsequently subtracted from the respective non-filtered phase image.  In order to normalise with respect to the incident beam, all amplitude images are divided by the average value inside the hole, whereas an offset phase found within the hole is subtracted from all phase images. Once images are normalised, they are aligned with respect to each other with a sub-pixel image registration algorithm \cite{manuel_sub_pixel}.  For $A_{\mathrm{XMCD}}$ images, the logarithm of the amplitude images is taken and then the difference between RCP ($A\mathrm{_{RCP}}$) and LCP ($A\mathrm{_{LCP}}$) are taken to subtract the electronic scattering factor. Similarly for the $\mathrm{\phi_{XMCD}}$ images, the difference between RCP and LCP is taken to subtract the electronic contribution to the scattering factor. The equations are as follows:
 \begin{equation*}
     A\mathrm{_{XMCD}} = \frac{\log_{e}(A_\mathrm{{RCP}}) - \log_{e}(A_\mathrm{{LCP}})}{2}
    \label{xmcd_def}
 \end{equation*}
 \begin{equation*}
      \mathrm{\phi_{\mathrm{XMCD}} = \frac{\phi_{RCP} - \phi_{LCP}}{2}}
 \end{equation*}
Quantitative data is extracted from the projections as follows.
The transmission as a function of energy is obtained by taking the square of the absolute value of the complex images taken at each energy, averaging over both positive and negative domains (plotted as a black line in Figure\,\ref{spectro_ptycho}). We observe two resonance peaks corresponding to the L$_3$ and L$_2$ absorption edges of Cobalt. In order to quantitatively extract the complex XMCD spectra from the images, we average the XMCD signal within each domain and obtain the spectra shown in Fig \ref{spectro_ptycho}b for both $A_{\mathrm{XMCD}}$ (red curve) and $\mathrm{\phi_{XMCD}}$ (blue curve). This is done by generating a boolean mask for the domains by running a Chan Vese segmentation algorithm \cite{chan_vese_segmentation} on a high Signal to Noise ratio $A\mathrm{_{XMCD}}$ image taken on resonance at 780\,eV. This gives two masks, one for each domain. These masks are then applied to the whole stack of aligned images to extract the magnetic signal from the same regions. The area taken from each image at each energy is then averaged to produce the spectra shown in Figure\,\ref{spectro_ptycho}b.

\section{Spectro-ptychography on FeGd samples}
\label{FeGd_1000nm_spectro_ptycho}
\begin{figure}[htbp]
  \centering
  \includegraphics[width=.5\textwidth, keepaspectratio=True]{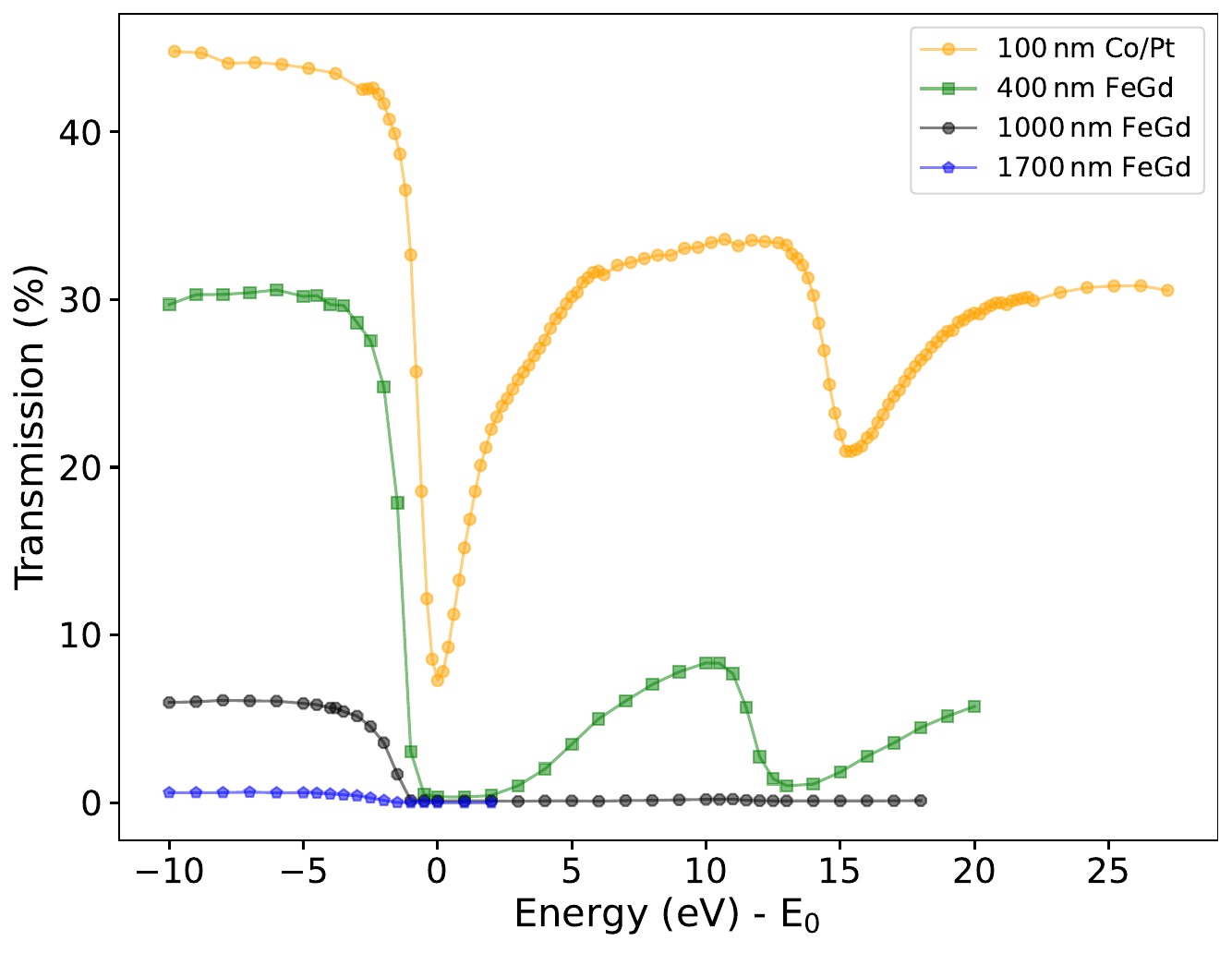}
  \caption{ Transmission signal through the thickness of the 100\,nm thick CoPt multi-layer (Yellow), 400 nm (green), 1\,$\mu$m (black) and 1.7\,$\mu$m (blue) FeGd samples. The energies are normalised with respect to the L$_3$ absorption edge of the materials at 780 eV for Co and 710 eV for Fe. Transmission only measurable at pre absorption edges for the thick FeGd samples. On resonance, the transmission is zero.}
  \label{FeGd_transmission}
\end{figure}

\begin{figure*}[htbp]
  \centering
  \includegraphics[width=.7\textwidth, keepaspectratio=True]{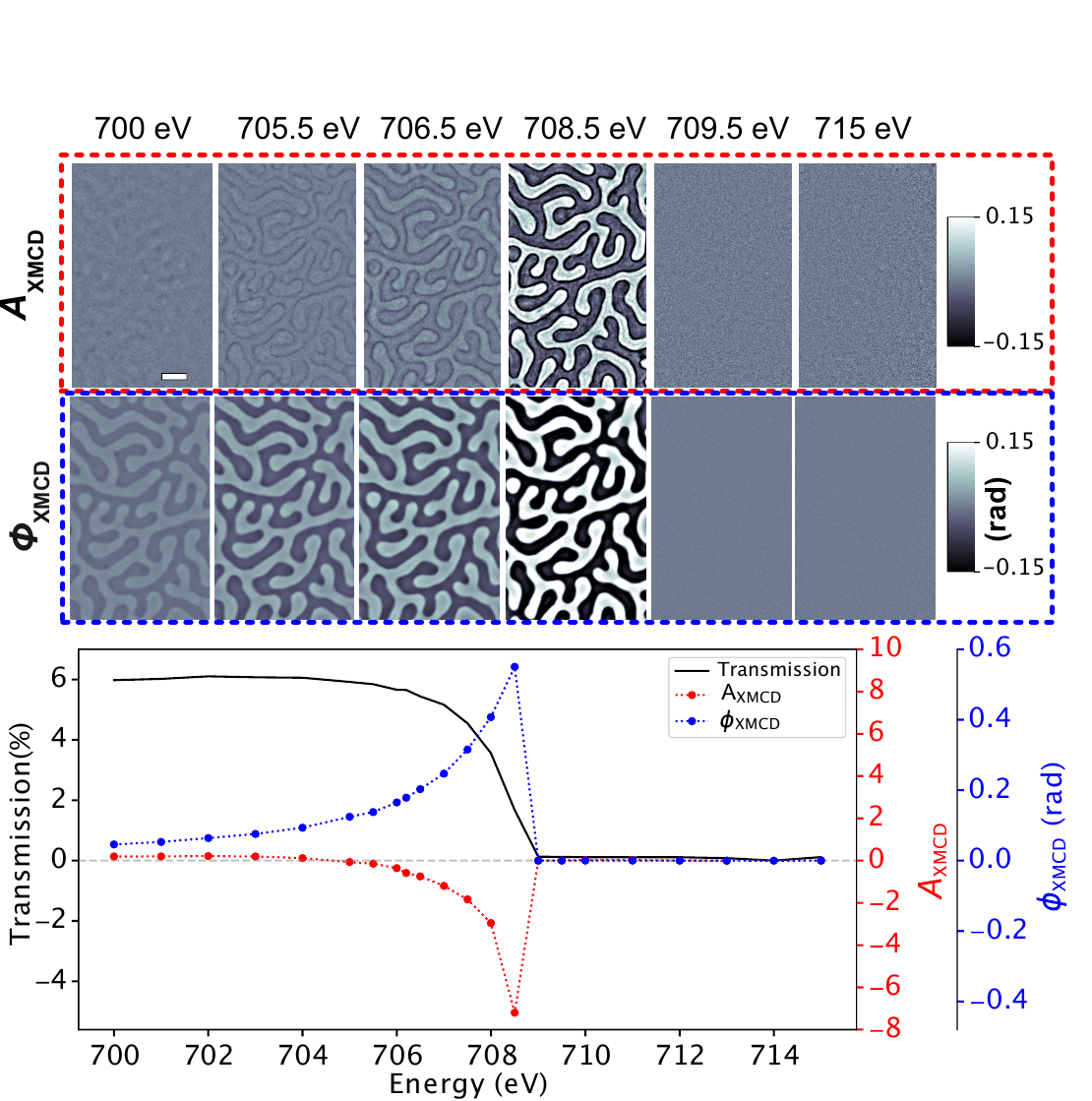}
  \caption{Pre-edge phase ptychography on 400\,nm , 1\,$\mu$m and 1.7\,$\mu$m thick FeGd samples close to the Fe L$_3$ edge. Complex XMCD spectrum across the Fe L$_3$ edge  extracted from magnetic domains in XMCD projections of a  1\,$\mu$m thick FeGd film with $A_{\mathrm{XMCD}}$ and $\phi_{\mathrm{XMCD}}$ shown in red and blue dots, respectively. The black line indicates the transmission through the sample. The images above the plot show $A_{\mathrm{XMCD}}$ and $\phi_{\mathrm{XMCD}}$ images for a few selection of energies. Scale bar represents 500\,nm.}
  \label{1000nm_spectro_ptycho}
\end{figure*}

We performed spectro-ptychography scans on FeGd samples of thicknesses 400\,nm, 1\,$\mu$m and 1.7\,$\mu$m across the the L$_3$ and L$_2$ edges of Fe. The transmission of the samples, plotted in Fig.\,\ref{FeGd_transmission}, decreases dramatically on resonance, with no transmission at the Fe L$_3$ absorption edge (normalised to zero Energy) for all FeGd samples. However, even for the 1.7\,$\mu$m thick film, there remains detectable transmission in the pre-edge.
By exploiting the $\phi_{\mathrm{XMCD}}$ in the pre-absorption edge, we were successfully able to image FeGd samples upto 1.7\,$\mu$m with high contrast and nanoscale spatial resolution. The complex XMCD spectra, along with a selection of $A_{\mathrm{XMCD}}$ and $\phi_{\mathrm{XMCD}}$ projections, are given for the 1\,$\mu$m thick film shown in Fig.\,\ref{1000nm_spectro_ptycho}.

\section{Simulation of thicker samples}
\label{simulation_thick_systems}
\begin{figure}[htbp]
  \centering
  \includegraphics[width=.5\textwidth, keepaspectratio=True]{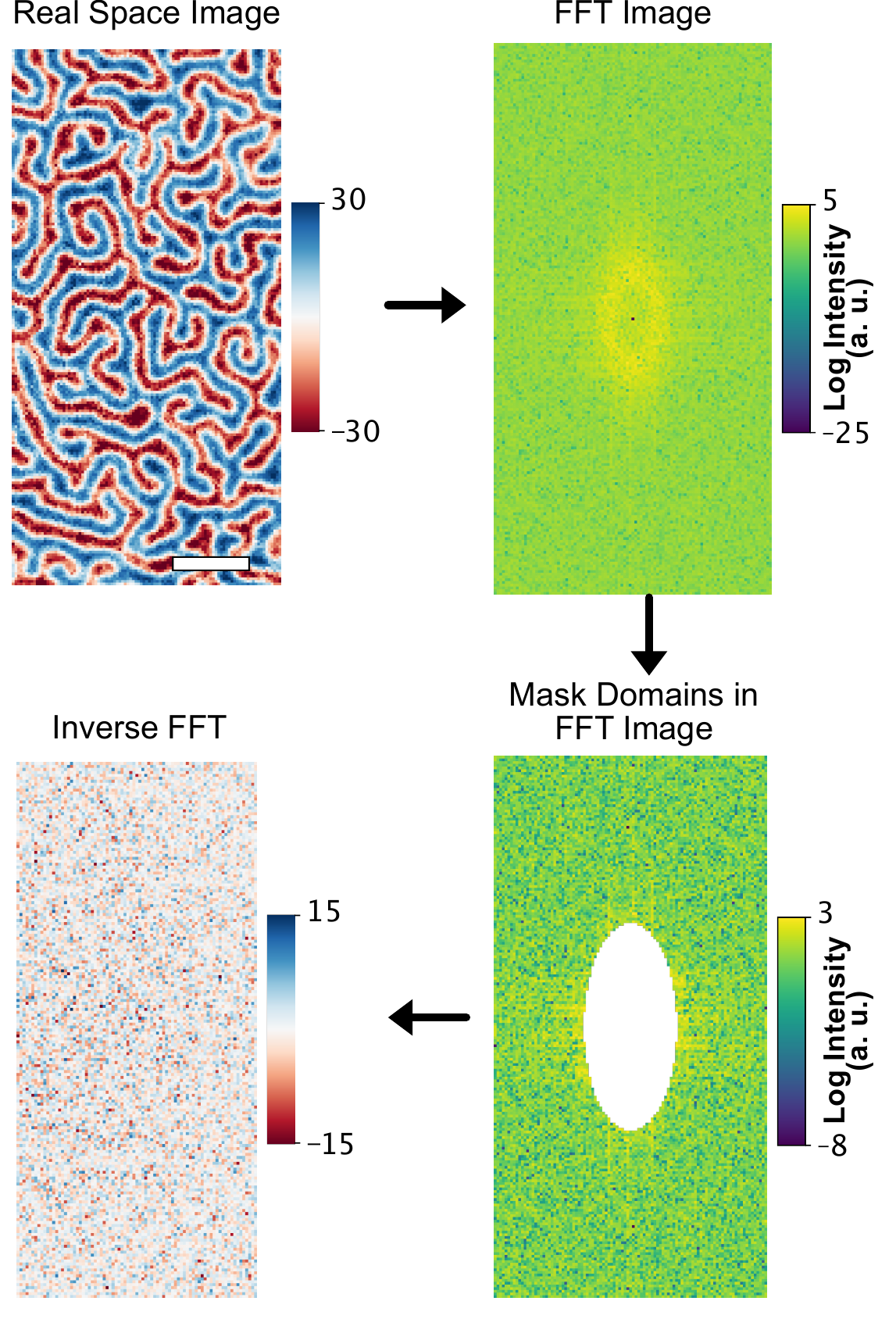}
  \caption{Protocol used for extracting noise in measurement data. Scale bar in the image represents 500\,nm.}
  \label{noise_protocol}
\end{figure}
To determine the signal to noise ratio (SNR) of the $A_{\mathrm{XMCD}}$ and $\phi_{\mathrm{XMCD}}$ images as a function of thickness, we separately determine the thickness evolution of the signal, noise and then the calculation of the SNR.

\subsubsection{Signal}
\label{Signal}
In order to simulate the transmission intensity for thicker samples, we calculate the absorption coefficient of the material with the measured transmission spectra (shown in Fig. \ref{spectro_ptycho}), from $I = e^{-\mu_\mathrm{{abs}} z}$, where I is the transmitted intensity through the sample ($I=A(E)^2$ with $A(E)$ being the measured amplitude), $\mathrm{\mu_{abs}}$ is the absorption coefficient and $z$ is the thickness of the material. Now, using this same equation, we can calculate the transmitted intensity as a function of energy for different thicknesses of a particular material, as shown in Fig.\,\ref{signal_vs_thickness}a.

To simulate the $A_{\mathrm{XMCD}}$ spectra for different thicknesses, we first calculate the transmission intensity ($I\mathrm{_{RCP(LCP)}}$) from the dichroic spectro-ptychography projections taken with, both RCP and LCP X-rays. We then separately obtain the absorption coefficients $\mathrm{\mu_{RCP(LCP)}} \propto \,\Im (f(E, r))$, assuming the magnetisation vector is oriented (anti)parallel to the direction of propagation of X-rays, similar to the measurements. Now using $I_\mathrm{{RCP(LCP)}} = e^{-\mu_\mathrm{{RCP(LCP)}} z}$ we calculate the transmission for various thickness for RCP ($\mathrm{I_{RCP}}$) and LCP ($\mathrm{I_{LCP}}$) X-rays. To isolate the A$_{\mathrm{XMCD}}$ signal, we use the following equation:
\begin{equation}
     A\mathrm{_{XMCD}} = \frac{\mathrm{log}_{e}(I_\mathrm{{RCP}}) - \mathrm{log}_{e}(I_\mathrm{{LCP}})}{4}
    \label{xmcd_def_transmission}
 \end{equation}
where the factor of 4 accounts for the transformation from intensity to amplitude. Additionally, to impose the constraint of transmission dropping to zero after a certain thickness, we set the value to zero below a certain threshold transmission of 0.05\,\% of the total transmission.

Similarly given that $\phi = \delta z$, with $\delta \propto \Re (f(E, r))$, assuming the magnetisation vector is oriented (anti)parallel to the direction of propagation of X-rays, similar to the measurements, we calculate $\delta$ from the measured dichroic phase spectra taken with RCP ($\mathrm{\phi_{RCP}}$) and LCP ($\mathrm{\phi_{LCP}}$) X-rays. Using the same equation, we then simulate the $\mathrm{\phi_{RCP}}$ and $\mathrm{\phi_{LCP}}$ phase spectra for thicker samples and the $\phi_{\mathrm{XMCD}}$ obtained for each thickness using the following equation.
 \begin{equation*}
      \mathrm{\phi_{\mathrm{XMCD}} = \frac{\phi_{RCP} - \phi_{LCP}}{2}}
 \end{equation*}

\subsubsection{Noise}
\label{Noise}
\begin{figure}[htbp]
  \centering
  \includegraphics[width=.4\textwidth, keepaspectratio=True]{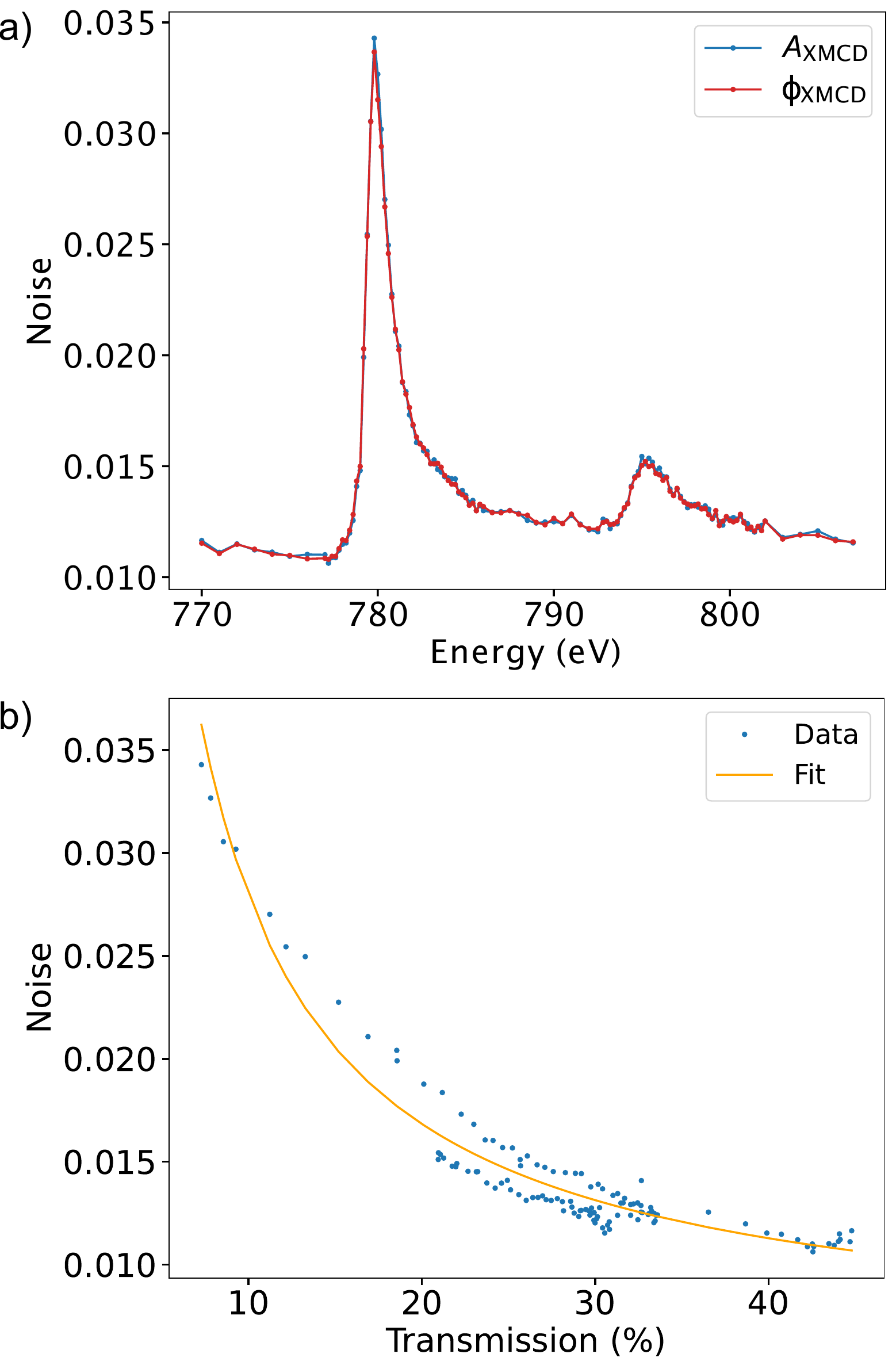}
  \caption{a)Noise as a function of energy for 100\,nm CoPt.The noise increases as we approach resonance as transmission decreases b) Noise as a function of transmission. The fit has a 1/x dependence which is used to calculate the noise for lower transmission corresponding to transmission through thicker samples.}
  \label{noise_figure}
\end{figure}

The noise is calculated differently for the 'thin' and 'thick regime'.
We first consider the 'thin' regime where the noise of the images for the 100\,nm CoPt and 400\,nm FeGd was calculated by taking the standard deviation of the high frequency noise present in the system. This was done by performing a Fast Fourier Transform (FFT) of an XMCD image, and masking the frequency regime corresponding to the domains. We then perform an inverse FFT and take the standard deviation of the filtered image. An example shown in Fig. \ref{noise_protocol}. The value of the noise as a function of energy is given in Fig \ref{noise_figure}a: as the transmission decreases, the noise increases. It can be seen that although the $A_{\mathrm{XMCD}}$ (dimensionaless) and $\phi_{\mathrm{XMCD}}$ (rad) are two different quantitites with, the noise as a function of energy is very similar. This is because both quantities are retrieved from the same diffraction patterns that are measured by the detector, and thus, the noise of the ptychography projections can be directly related to the intensity on the detector.   

% \begin{figure}[htbp]
%   \centering
%   \includegraphics[width=.4\textwidth, keepaspectratio=True]{1000_nm_noise_images.pdf}
%   \caption{a)$A_{XMCD}$ and b) $\phi_{XMCD}$ images showing low frequency noise components in white due to the vanishing transmission seen in a few regions in the domain. }
%   \label{1000nm_images_noise}
% \end{figure}

For the 'thick regime' with the 1\,$\mu$m and 1.7\,$\mu$m thick FeGd , the noise also includes modulated noise within the domains, due to contrast only present in the vicinity of the domain walls. Hence, in order to calculate the noise for the images of 1\,$\mu$m and 1.7\,$\mu$m thick FeGd, the average of the standard deviation of the XMCD signal seen within each domain gives the A$_{\mathrm{XMCD}}$ and $\phi_{\mathrm{XMCD}}$ noise in the images.

\subsubsection{Calculation of SNR}
\label{SNR_calculation}
Having obtained the noise from the images, we calculate the dimensionless SNR for each sample separately using the respective signal obtained from measurements. The maximum SNR value for each of the samples is then taken and plotted as a function of thickness for all measured samples, shown in Fig. \ref{thick_imaging}b as red and blue dots for the A$_{\mathrm{XMCD}}$ and $\phi_{\mathrm{XMCD}}$ respectively.

\begin{figure}[htbp]
  \centering
  \includegraphics[width=.5\textwidth, keepaspectratio=True]{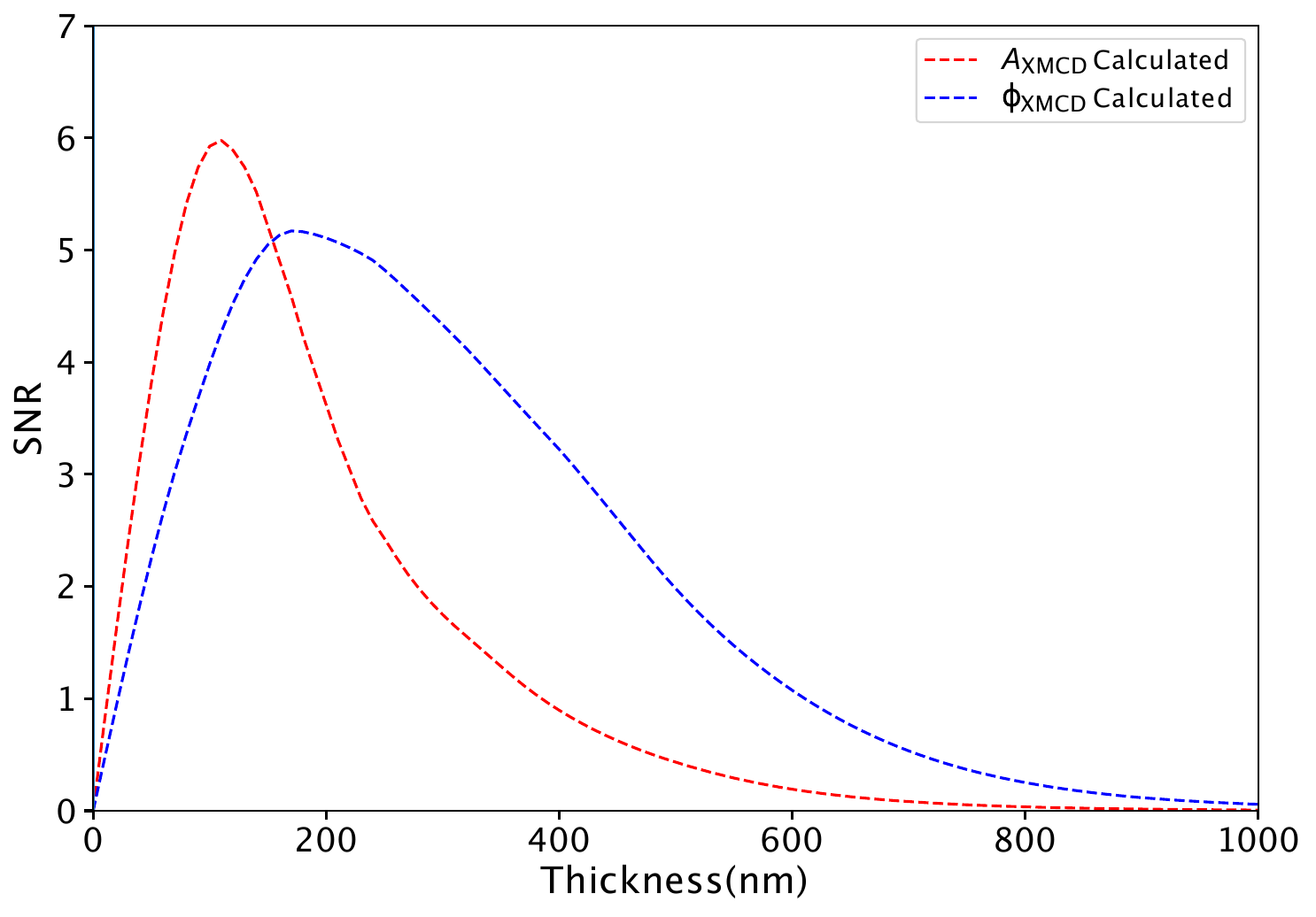}
  \caption{Simulated SNR of $\mathrm{A_{XMCD}}$ and $\mathrm{\phi_{XMCD}}$. The $\mathrm{\phi_{XMCD}}$ SNR is higher for increasing thickness than the $A\mathrm{_{XMCD}}$. This plot is scaled in thickness, for a direct comparison with experimental data, shown in Fig. \ref{thick_imaging}b.}
  \label{CoPt_snr_thickness}
\end{figure}

To determine the noise for the simulated data, we first obtain a relation between the measured noise and transmission, by plotting the noise as a function of transmission in Fig. \ref{noise_figure}b, where we see a $\frac{1}{x}$ dependence in the noise. Using this relation we  calculate the noise we expect to see for thicker samples and calculate the SNR for the simulated data for CoPt, which is plotted as dashed lines in Figure\,\ref{CoPt_snr_thickness}.

In order to compare the simulated CoPt SNR curves with measured FeGd we arbitrarily scale the thickness of both simulated SNR curves to match the experimental data as shown in Fig. \ref{thick_imaging}b.
This calculation of the dimensionless SNR allows us to compare the quality of the images originating from the two contrast mechanisms.

The thickness of 100\,nm thick CoPt was matched to an effective thickness of 217\,nm thick FeGd by comparing and scaling the measured transmission spectra between 100\,nm thick CoPt 400\,nm thick FeGd. The main difference between the two arises from the different scattering factor, i.e. absorption coefficient $\mu$. Although resonance occurs at different energies, by comparing both measured transmission  spectra, we can estimate a scaling factor of 1.45 to effectively match the transmission of the two samples to yield an effective thickness with respect to each other.

\section{Spatial resolution}
\label{FRC_spatial_res}

\begin{figure}[htbp]
  \centering
  \includegraphics[width=.49\textwidth, keepaspectratio=True]{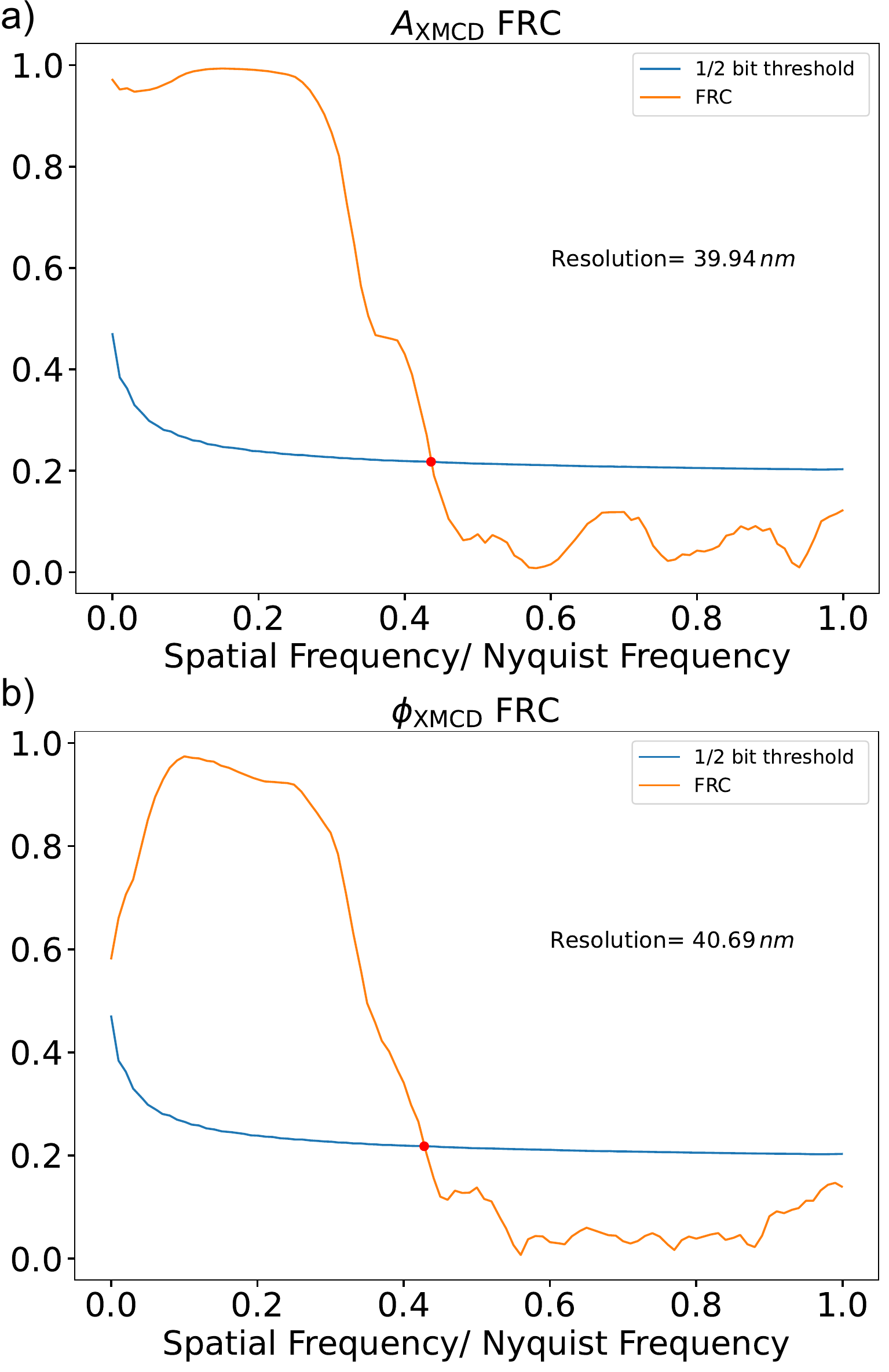}
  \caption{Fourier Ring Correlation with half-bit threshold for 100\,nm thick CoPt on a) $A_{\mathrm{XMCD}}$ image taken at 780\,eV and b) $\phi_{\mathrm{XMCD}}$ image taken at 779.4\,eV.}
  \label{spatial_res}
\end{figure}

The full-period spatial resolution of the images with the strongest XMCD contrast are calculated by Fourier Ring Correlation (FRC) with a half bit threshold \cite{FRC}. An example is shown in Fig \ref{spatial_res}, and the values observed for the different thickness are summarised in the Table \ref{FRC}. \\
In comparison, the $\phi_{\mathrm{XMCD}}$ in general exhibits a higher spatial resolution calculated with the FRC in comparison with the $A_{\mathrm{XMCD}}$. Although ultra high nanoscale spatial resolutions are possible to measure with Ptychography, they are not achieved here as it was not the goal of the experiment. Rather than devoting the statistics to high spatial resolution, the limited beamtime awarded was devoted instead to measuring the various samples with high energy resolution. In order to obtain higher spatial resolutions, the experiment could be further optimised, acquiring a larger portion of the detector, and acquiring higher statistics  to further reduce the noise in the images.
\\
\begin{table}[htbp]
    \centering
    \begin{tabular}{ |p{2.5cm}|p{1.8cm}|p{1.2cm}|  }%{|c|c|c|c|c|}
        \hline
        \multirow{2}{*}{Thickness (nm)} &\multicolumn{2}{c|}{FRC (nm)}\\
          & Amplitude & Phase \\
        \hline
         100 CoPt& 39.94 & 40.69 \\
         400 FeGd& 54.68 & 47.70 \\
         1000 FeGd& 52.13 & 47.20 \\
         1700 FeGd& 57.32 & 56.04\\
         \hline
    \end{tabular}
    \caption{Summary of the spatial resolution by Fourier Ring Correlation (FRC) with a half bit threshold. The phase has a higher spatial resolution than the amplitude for thicker samples.}
    \label{FRC}
\end{table}

\end{document}